\documentclass[twocolumn,aps,prb]{revtex4-1}
\usepackage{natbib}
\usepackage{amsmath}
\usepackage{amssymb}
\usepackage{amsthm}
\usepackage{epstopdf}
\usepackage{bm}
\usepackage{hyperref}
\usepackage{epsfig}
\usepackage{dcolumn}
\usepackage{url}

\def\sgn{\mathrm{sgn}}

\begin{document}

\title{Observing Majorana bound states of Josephson vortices in topological superconductors}

\author{Eytan Grosfeld$^1$ and Ady Stern$^2$}
\affiliation{$^1$Department of Physics, University of Illinois,
1110 W. Green St., Urbana IL 61801-3080, U.S.A. \\
$^2$Department of Condensed Matter Physics, Weizmann Institute of Science, Rehovot 76100, Israel}

\begin{abstract}
In recent years there has been an intensive search for Majorana fermion states in condensed matter systems. Predicted to be localized on cores of vortices in certain non-conventional superconductors, their presence is known to render the exchange statistics of bulk vortices non-Abelian. Here we study the equations governing the dynamics of phase solitons (fluxons) in a Josephson junction in a topological superconductor. We show that the fluxon will bind a localized zero energy Majorana mode and will consequently behave as a non-Abelian anyon. The low mass of the fluxon, as well as its experimentally observed quantum mechanical wave-like nature, will make it a suitable candidate for vortex interferometry experiments demonstrating non-Abelian statistics. We suggest two experiments that may reveal the presence of the zero mode carried by the fluxon. Specific experimental realizations will be discussed as well.
\end{abstract}

\maketitle


\section{Introduction}

Non-Abelian statistics \cite{MooreRead, NayakWilczek,RMP} has recently been the subject of intensive research driven both by its possibly profound impact on the field of quantum computation \cite{Kitaev, DasSarmaNayakFreedman,FreedmanLarsenWang} and by the search for its manifestations \cite{SternHalperin, BondersonKitaevShtengel}. Among all mechanisms giving rise to such statistics, the route via spin-polarized p-wave superfluidity may be the simplest one. It was previously argued \cite{KopninSalomaa,ReadGreen,Ivanov} that an Abrikosov vortex in a $p$-wave superfluid can trap a zero energy Majorana fermion, being a self-conjugate ``half'' fermion. A pair of Majorana modes constitute a regular fermion, and the resulting non-local occupancies label a set of degenerate ground states. Braiding of vortices results in mixing of these ground states, sometimes in a non-commutative fashion: it matters in which order multiple braidings are performed. The search for an explicit experimental signal of the resulting vortex exchange statistics, as well as for the presence of Majorana modes on their cores, is currently on its way.

In this paper we propose an experiment that probes Majorana fermions in Josephson vortices (fluxons). Josephson vortices are trapped in insulating regions between super-conductors. For conventional superconductors, they are described as solitonic solutions of the sine-Gordon equation moving with small inertial mass (estimated to be smaller than the electron mass). In the case of topological superconductors, we find that such vortices bind a localized Majorana zero mode and would therefore behave as non-Abelian anyons, despite the fact that they lack a normal core. We show that the non-abelian nature of these vortices manifests itself in measurable transport properties of the Josephson junctions that house them.

\begin{figure}[htp]
\centering
\includegraphics[scale=0.32]{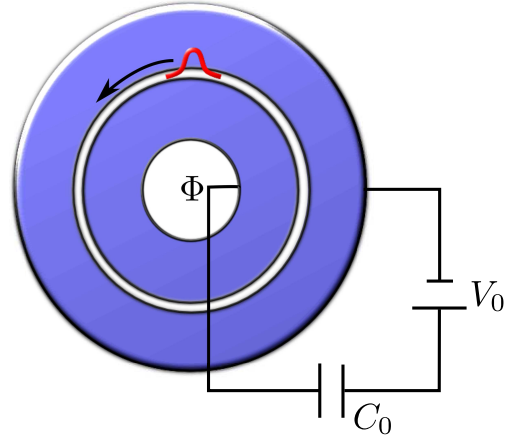}
\caption{Aharonov-Casher effect in a long circular Josephson junction. The junction traps a single fluxon which is traveling around the ring propelled by a bias charge $Q$ induced between the two ring-shaped superconductors. The energy spectrum of the junction is periodic in $Q$ with periodicity $e$ when $\Phi$ is increased to nucleate a vortex within the interior hole.
Copper wires act as reservoirs of unpaired electrons.}
\label{fig:long-junc}
\end{figure}

We start by showing that Josephson vortices  traveling in a Josephson junction in a topological superconductor bind a single Majorana zero energy state (see Equation \ref{eq:majorana}). We then  discuss two experiments that can be used to measure the presence of these Majorana fermions. The first probes a thermodynamical property of a circular charge biased Jospephson junction (Figure \ref{fig:long-junc}), by measuring the non-linear capacitance induced by the persistent motion of the vortex trapped in the junction\cite{ReznikAharonov,Wees, HermonSternBenJacob,Elion}. The second is an interference experiment of fluxons demonstrating Aharonov-Casher \cite{AharonovCasher} oscillations (Figure \ref{fig:interferometer}), similar in spirit to the one proposed in \cite{GrosfeldSeradjehVishveshwara}.

\section{Hamiltonian of a circular Josephson junction}

We start by considering a circular Josephson junction, made of two concentric super-conducting annuli, separated by a thin insulator.
We assume that the hole at the center of the inner super-conductor is of a size comparable to the superconducting coherence length, and encloses $N_v$ vortices. The Hamiltonian governing the junction would be composed of three parts, $H=H^\phi+H^\psi+H_{\mathrm{tun}}$. The first, $H^\phi$ is related to the dynamics of the phase across the junction. For a Josephson junction of height $h_z$, this part of the dynamics is derived from the following Hamiltonian (see e.g. \cite{Tinkham, HermonSternBenJacob})


\begin{eqnarray}
	\nonumber H^\phi&=&\hbar \bar{c}\int_x\left\{\beta^2 \frac{h_z^2}{2}(n-\sigma)^2\right.\\
	&&\left.+\frac{1}{\beta^2}\left[\frac{1}{2}(\partial_x\phi)^2+\frac{1}{\lambda^2_J}(1-\cos\phi)\right]\right\},
\label{eq:phase-hamiltonian}
\end{eqnarray}
where $\phi$ is the phase difference across the junction, $n$ is the two dimensional (2D) density of Cooper pairs on, say, the inner plate, and $\sigma$ is the 2D density of the externally induced charge. The first part of the Hamiltonian is the capacitive energy, the second the magnetic energy, and the third is the Josephson energy. The resulting equation of motion is the sine-Gordon equation, with the typical speed of light reduced to $\bar{c}^2=c^2\frac{d}{d+2\lambda_L}$ (here $d$ is the width of the insulating barrier and $\lambda_L$ is the London penetration length). The Josephson vortex is a soliton described by this equation, with its typical size set by the Josephson penetration length $\lambda_J$. 
We shall assume throughout that the circumference of the junction, $L$, is much larger than $\lambda_J$, and that $\lambda_L \ll h_z\ll \lambda_J$. The parameter
$\beta^2\equiv 16\pi\frac{e^2}{\hbar c}\sqrt{d(2\lambda_L+d)}/h_z$.

The second part of the Hamiltonian, $H^\psi$, originates from the neutral protected edge modes of the topological superconductor, which give rise to its quantized thermal Hall conductance. The Hamiltonian governing these neutral modes is $H^\psi=H^\psi_1+H^\psi_2$ with
\begin{eqnarray}
	\label{eq:neutral} H^\psi_{1,2} = \pm i v_\psi \int dx\,\psi_{1,2}(x)\partial_x\psi_{1,2}(x),
\end{eqnarray}
where $1$ and $2$ refer to the two counter-propagating Majorana edge modes. Here $\psi_{i}(x)$ is a Majorana field, $\psi^\dag_i(x)=\psi_i(x)$, $x$ is the coordinate running along the Josephson junction, and $v_\psi$ is the velocity along the edge. In terms of the electron's creation and annihilation operators $c_i(x),c_i^\dagger(x)$ and the phase fields of the two super-conductors on the two sides of the junctions $\phi_i(x)$ the Majorana fields may be expressed as
\begin{eqnarray}
	\psi_i(x)=e^{i\phi_i(x)/2-i \pi x/L}c_i(x)+e^{-i\phi_i(x)/2+i \pi x/L}c_i^\dag(x),
\end{eqnarray}
In this explicit form, the equation holds true for a spin-polarized $p$-wave superconductor. For a general topological superconductor, it is still the case that the Majorana field will acquire a minus
sign going around the edge, in addition to one minus sign per each
vortex enclosed in its path (i.e., one minus sign for every $2\pi$ winding of the phase $\phi_i$). The boundary conditions of the fields $\psi_1(x),\psi_2(x)$ therefore depend on $N_v$, the number of vortices enclosed by the two super-conducting annuli.

When the two superconducting islands are brought into  close proximity, tunneling terms of the form $c_1^\dagger(x)c_2(x)$ translate to
\begin{eqnarray}
	\label{eq:tunneling} H_{\mathrm{tun}} &=& 2 i m\int dx\, \psi_1(x)\psi_2(x)\cos(\phi/2),
\end{eqnarray}
with $\phi(x)\equiv\phi_2(x)-\phi_1(x)$ and $m$ being a tunneling amplitude. Writing Equations (\ref{eq:neutral}) and (\ref{eq:tunneling}) compactly as a matrix equation, the Hamiltonian $H^\psi_1+H^\psi_2+H_{\rm tun}$ becomes
\begin{eqnarray}
	\nonumber & H=\int dx\, \Psi^T(x)\left[\begin{array}{cc}
	i v_\psi \partial_x & i m\cos\left(\phi/2\right) \\
	-i m\cos\left(\phi/2\right) & -i v_\psi\partial_x
	\end{array}\right]\Psi(x),\\
	\label{eq:neutral-hamiltonian}
\end{eqnarray}
where $\Psi=(\psi_1,\psi_2)^T$ is a spinor composed of the two counter propagating Majorana modes. The Hamiltonian possesses a symmetry under $\phi\to\phi+2\pi$ and $\Psi\to\sigma_z \Psi$.


\section{Bound Majorana mode on the background of a soliton}

A solitonic solution of Equation (\ref{eq:phase-hamiltonian}), also known as a fluxon or a Josephson vortex, is a finite energy solution which interpolates between two minima of the periodic potential described by the Josephson term. For a long Josephson junction ($L\gg \lambda_J$) it acquires the form $\phi_s(x)=4 \arctan\left[\exp\left(\frac{x-x_0}{\lambda_J}\right)\right]$ where $x_0$ is the position of the soliton (see e.g. \cite{Tinkham, HermonSternBenJacob}). In the following we solve Equation (\ref{eq:neutral-hamiltonian}) in the background of a single soliton, explicitly plugging $\phi_s$ into $\phi$, and using $\cos\left(\frac{\phi_s(x)}{2}\right)=-\tanh\left(\frac{x-x_0}{\lambda_J}\right)$. This would in turn result in a tunneling amplitude $m\cos\left({\frac{\phi_s}{2}}\right)$ whose sign is different on the two sides of the soliton. In light of the Jackiw-Rebbi mechanism, Equation (\ref{eq:neutral-hamiltonian}) will now bind a zero energy mode at the position of the soliton $x_0$,
\begin{eqnarray}
	\label{eq:majorana} \gamma_J=\int dx\, f(x) \left[\psi_1(x)+\sgn(m)\psi_2(x)\right].
\end{eqnarray}
In the limit of a long junction, $L\gg \lambda_J$ and $L\gg v_\psi/m$,  the shape of the Majorana mode is described by the localized function
$f(x)=\frac{1}{N}\left[\cosh\left({\frac{x-x_0}{\lambda_J}}\right)\right]^{-|m|\lambda_J/v_\psi}$, with
$N$ a normalization factor and $x_0$ the center of the Josephson vortex. The operator $\gamma_J$ is a \emph{localized} Majorana fermion and the subscript $J$ indicates that this mode is bound to a Josephson vortex. This mode satisfies  $\gamma_J^\dag=\gamma_J$. Indeed, the entire low energy spectrum of bound states can be extracted. Plugging the solitonic solution into Equation (\ref{eq:neutral-hamiltonian}) and rotating the spinors according to $\Psi\to\frac{1}{\sqrt{2}}\left(\begin{array}{cc} 1 & 1 \\ -1 & 1\end{array}\right)\Psi$ the Hamiltonian can be written conveniently as
\begin{eqnarray}
	\label{eq:super} H=\left(\begin{array}{cc}
	0 & A^\dag \\
	A & 0
	\end{array}\right), \;\;\mbox{with}\;\; \left\{\begin{array}{l}A=-i v_\psi\partial_x-i W(x) \\ A^\dag=-i v_\psi\partial_x+i W(x)\end{array}\right.
\end{eqnarray}
The spectrum of this Hamiltonian possesses a supersymmetry with a superpotential $W(x)=m\tanh\left(\frac{x}{\lambda_J}\right)$. In particular, the zero mode is annihilated by either $A$ or $A^\dag$ depending on the sign of the mass, while the other operator will not have a normalisable eigenfunction at zero energy. The rest of spectrum is doubly degenerate with $A$ and $A^\dag$ connecting the states of the doublet. Squaring the Hamiltonian in Equation (\ref{eq:super}) and linearizing the potential we get a shifted harmonic oscillator, and the spectrum of excitations above the zero energy mode is
\begin{eqnarray}
	\label{eq:spectrum-soliton} E_n\sim \sqrt{\frac{\hbar v_\psi m}{\lambda_J}} \sqrt{2n}, \;\;\; \mbox{with} \;\;\; n=0,1,2,\ldots
\end{eqnarray}

Majorana fermions have to come in pairs, due to the properties of the BdG equations. The second Majorana fermion will be on one of the super-conductor edges, and will be denoted by $\gamma_e$, with $e$ standing for edge. Its position will depend on the number of vortices $N_v$ in the inner hole. A zero mode is found on every edge that encloses an odd number of vortices. Hence, when $N_v$ is odd, $\gamma_e$ will be localized on the edge separating the inner super-conductor from the vacuum. When $N_v$ is even, $\gamma_e$ will be localized on the edge separating the outer super-conductor from the vacuum. The two Majorana fermions, $\gamma_e$ and $\gamma_J$, cannot shift from zero energy without hybridizing, and the super-conductor that separates them prevents that from happening. Note that in the absence of electron tunneling across the junction, the presence of these two Majorana modes is protected by an index theorem. Their wavefunction will be spread evenly around the two edges of the superconductor. When we next turn on electron tunneling, the Majorana mode $\gamma_J$ cannot disappear without hybridizing with the distant Majorana mode on the edge of the sample. Therefore it necessarily persists in the junction. Due to the effect of tunneling it gets localized around the center of the soliton as described by equation (\ref{eq:majorana}).

Due to the presence of the soliton in the junction, the boundary conditions are different for the two Majorana fields and depend on $N_v$: $\psi_1(x)=-(-1)^{N_v}\psi_1(x+L)$ and $\psi_2(x)=(-1)^{N_v}\psi_2(x+L)$. A rotation of the soliton around the junction shifts $x$ by $L$ and $\phi_1(x)$ by $2\pi$. These shifts have two effects. First, they multiply $\psi_1$ by $-(-1)^{N_v}$ and $\psi_2$ by $(-1)^{N_v}$ due to the boundary conditions. Second, they multiply the off-diagonal term of (\ref{eq:neutral-hamiltonian}) by $-1$ and hence multiply $\psi_1(x)$ in Equation (\ref{eq:majorana}) by $-1$. Combining these two effects together, we see that $\gamma_J$ is multiplied by $(-1)^{N_v}$. In view of this, the transformation $\gamma_J\rightarrow(-1)^{N_v}\gamma_J$ may be understood as a consequence of the winding of one Majorana fermion around another for odd $N_v$, and the absence of such winding for even $N_v$. Remarkably, we find below that the capacitance of the junction in the two cases is different.


\section{Manifestation of the Majorana mode in thermodynamics and transport}

We now consider the energy of the long circular Josephson junction when it is biased by an external charge $Q=2e\sigma L h_z$. For a ``conventional'' Josephson junction made of $s$-wave super-conductors this problem was studied in \cite{HermonSternBenJacob}. In that case the Hamiltonian consists of Equation (\ref{eq:phase-hamiltonian}) only. The magnetic and Josephson energies constitute the rest mass of the vortex. The charging energy constitutes its kinetic energy, and is the only component that depends on $Q$. This kinetic/charging energy may be understood in two ways. First, by viewing the vortex as a particle of mass $M=\frac{\hbar^2 h_z}{2\pi e^2 d \lambda_J}$ in a one dimensional ring, subjected to a vector potential $\frac{Q}{2e}\frac{2\pi\hbar}{L}$, with a set of energy eigenstates
\begin{eqnarray}
E_{n}(Q)=\frac{(2\pi\hbar)^2}{2ML^2}\left(n-\frac{Q}{2e}\right)^2,
\label{vortex-kinetic-energy}
\end{eqnarray}
with $n$ being the quantum number that quantifies the momentum of the vortex. Second, by writing this energy as a capacitive energy of the form,
$E_{n}(Q)=\frac{(2e)^2}{2C}\left(\frac{Q}{2e}-n\right)^2$ where now $C$ is the effective capacitance (which for a short junction coincides with the geometric capacitance). The quantum number $n$ is now identified as the number of Cooper-pairs charging the Josepshon junction, and the capacitance is $C=M(2 e L)^2/(2\pi\hbar)^2$.

The spectrum of (\ref{vortex-kinetic-energy}) is described by a set of parabolas, each characterized by a different integer value of $n$. When $Q$ is increased so that the spectrum reaches a crossing point of two parabolas of $n$ and $n+1$, it becomes energetically favorable to tunnel a Cooper pair to decrease the charging energy, thus crossing to the next parabola. The matrix element required for this tunneling is generated by a small amount of disorder. The velocity of the moving fluxon is proportional to the derivative of the ground state energy $E_\mathrm{gs}(Q)=\min_{{n}\in {\mathbb Z}} E_{n}(Q)$ with respect to $Q$. In turn, the moving vortex generates a measurable voltage via the Josephson relation between the inner ring and outer ring of the device, which will be oscillating as function of $Q$ with a periodicity of $2e$. The periodicity of $2e$ is quite expected, given the energy cost needed to break a Cooper-pair. None of these results depend on the value of $N_v$.

The same spectrum shows a strikingly different behavior when the junction is made of two  topological superconductors. In particular, this difference manifests itself in the dependence of the spectrum on the charge $Q$ and $N_v$. The fermionic part of the Hamiltonian now has two degenerate eigenstates, which are two eigenstates of the operator $i\gamma_J\gamma_e$, with eigenvalues $\pm 1$. When it comes to the spectrum of the Josepshon vortex, there is a crucial difference between the case where $\gamma_e$ is localized at the edge of the outer super-conductor (even $N_v$) and that in which it is localized at the center of the inner super-conductor (odd $N_v$). As we saw before, in the latter case the winding of the soliton around the junction multiplies $\gamma_J$ by $-1$. It is easy to see that it multiplies $\gamma_e$ by $-1$ as well. These two operator transformations are implemented by the unitary transformation $U=e^{i\alpha}i\gamma_J\gamma_e$ on the state of the system \cite{NayakWilczek}. The phase $\alpha$ cannot be determined by this consideration, and is not important for what follows. The two ground states are eigenstates of $U$, and thus the application of $U$ multiplies them by two phases, that differ by $\pi$. The phase accumulated by the soliton encircling the ring affects its spectrum. When the energy (\ref{vortex-kinetic-energy}) is viewed as the energy of a charged particle on a ring threaded by a magnetic flux, a $\pi$ phase shift corresponds to the introduction of half a quantum of flux. The spectrum (\ref{vortex-kinetic-energy}) is then changed to be
\begin{eqnarray}
	E_{n,f}(Q)=\frac{(2e)^2}{2C}\left(\frac{Q'}{2e}-n-f\frac{1}{4}\right)^2,
\label{tsc-spectrum}
\end{eqnarray}
where the $f=\pm 1$ correspond to the two  eigenstates of $U$ for $\alpha=0$. The charge $Q'$ may be shifted with respect to the induced charge $Q$ by a non-universal number that depends on $\alpha$. The spectrum is now periodic with a period of $e$, rather than the period of $2e$ as was the case for the non-topological super-conductor, as well as for the topological super-conductor with an even $N_v$.

For the two sets of parabolas to be observed in an experiment when $N_v$ is odd, the system has to be able to switch between the two values of $f$, namely the two states $i\gamma_J\gamma_e=\pm 1$. This switch requires the total number of electrons to change by one \cite{SternHalperin,IlanGrosfeldStern,Liang}. To allow for that to happen, the junction should be weakly coupled to a reservoir of single electrons, i.e., to a metal.  The ground state energy would be $E_\mathrm{gs}(Q)=\min_{n\in {\mathbb Z},f\in {\mathbb Z}_2} E_{n,f}(Q)$. The end result is that the voltage measured between the inner and outer edges of the superconductor would be periodic as function of $Q$ (or $V_0$) with the periodicity now being $e$ for odd $N_v$ and $2e$ for even $N_v$.

\begin{figure}[htp]
\centering
\includegraphics[scale=0.32]{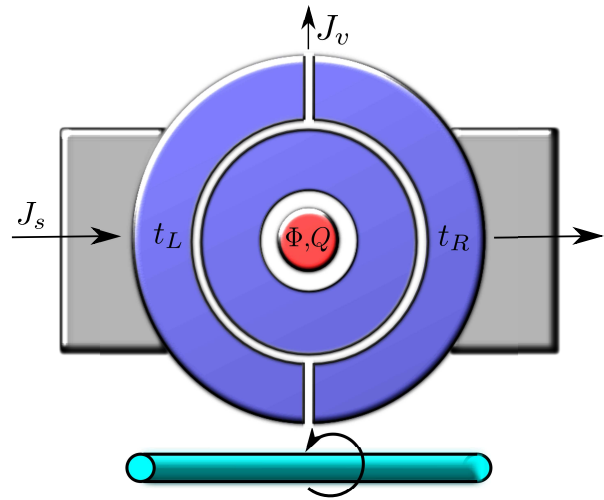}
\caption{Vortex interferometry experiment based on the Aharonov-Casher effect \cite{GrosfeldSeradjehVishveshwara} adapted to Josephson vortices. A superconducting wire creates a circulating magnetic field acting as a source for the entrance of Josephson vortices into the sample. An applied super-current drives the vortex along one of two paths circumventing an island towards the top of the sample. A charge $Q$ enclosed in the island controls the interference term via the Aharonov-Casher effect. When the flux $\Phi$ nucleates a vortex in the central region, the interference term would be obliterated.}
\label{fig:interferometer}
\end{figure}

While this experiment probes a thermodynamical property, the next one is an interference experiment that probes electronic transport. The fluxons are generated near the bottom of the interferometer (Figure \ref{fig:interferometer}), and driven by a super-current driven from left to right. The fluxon beam is split into two partial waves enclosing an island with externally imposed charge $Q$. Due to the Aharonov-Casher effect, the magnitude of the vortex current oscillates according to
\begin{eqnarray}
	J_v=J_{v0}\left[1+\zeta\cos\left(2\pi \frac{Q}{2e}\right)\right],
\end{eqnarray}
where $\zeta$ is the visibility of the oscillations and $J_{v0}$ the average current. By the Josephson relation, a measurable voltage difference is created between the left and right ends of the superconductor, which would oscillate with changing $Q$. If the flux $\Phi$ on the central island is increased so that another non-Abelian vortex is nucleated in the central hole, the interference pattern would be zero, $\zeta=0$. These results are sensitive to effects of decoherence, which would be reflected in a reduction of the value of $\zeta$ as the temperature is increased. However, the fluxons are quite light, and it was predicted \cite{ShnirmanBenJacobMalomed} and experimentally demonstrated \cite{Elion,Wallraff} that they can exhibit quantum behavior \cite{Hassler,ClarkeShtengel}. Also, the gap to the next fermionic state on the background of the soliton has different parametric dependence as compared to Abrikosov vortices.



For a soliton in a topological superconductor there are two main energy scales that control decoherence effects. The first is related to the bosonic part $H^\phi$, while the second stems from the fermionic part $H^\psi$.

The spectrum of the bosonic Hamiltonian $H^\phi$ contains non-topological excitations with which the fluxon can interact. These are quantized phase oscillations, or plasmons. Written in terms of collective coordinates for the fluxon, the Hamiltonian would be similar to the Caldeira-Leggett Hamiltonian of a quantum particle interacting with a bath. However, unlike the Caldeira-Leggett mechanism, here the plasmon spectrum is gapped, with the plasmon gap given by $\hbar\omega_p=\hbar\bar{c}/\lambda_J$.
At energy scales below this gap (estimated to be a few Kelvin) the coupling to plasmons is exponentially suppressed, and the internal dephasing thus mostly avoided \cite{HermonShnirmanBenJacob} over lengths much larger than the Josephson penetration length. This lies at the origin of the prediction of quantum behavior of Josephson vortices in long Josephson junctions. In particular, scattering of the fluxon on static inhomogeneities in the junction would be mostly elastic due to the presence of this plasmon gap\cite{ShnirmanBenJacobMalomed}. Tunneling of a fluxon through barriers was predicted to be enhanced by the presence of the plasmons \cite{ShnirmanBenJacobMalomed} and experimentally observed \cite{Wallraff}.

The second energy scale is unique to topological superconductors, and is related to the presence of higher energy fermionic states above the zero energy mode carried by the soliton, as described by $H^\psi$ and the fermionic tunneling terms, see Equation (\ref{eq:spectrum-soliton}). The presence of these states would result in decoherence of the non-Abelian mechanism. The gap to the first state would be $\hbar \omega_n=\sqrt{2\hbar v_\psi m/\lambda_J}$. To avoid these effects, the temperature should be lower than this gap as well.


\section{Proposed realizations}

There are several systems in which the ideas presented above may be implemented in practice. The most relevant for our purpose is a topological insulator whose surface states become superconducting by proximity to an $s$-wave superconductor, and are driven into an effective $p$-wave pairing for a single fermionic species. The second is a hybrid structure of three materials: a semiconductor with a magnetic material and a superconductor layer placed on top of it, as discussed in \cite{SauLutchynTewari}. The third is the perovskite material Sr$_2$RuO$_4$ (SRO) which becomes superconducting at temperatures below $1.5$K, and for which there is growing evidence that it realizes an unconventional pairing of a $p_x\pm i p_y$ (spin-triplet) form.

The surface state of a topological insulator is described by the Hamiltonian $H_0=v_F \psi^\dag\hat{z}\cdot \boldsymbol{\sigma}\times(-i\boldsymbol{\nabla})\psi$ (where $\psi=(\psi_\uparrow,\psi_\downarrow)^T$ is a spinor composed of the electronic operators associated with the spin components). The chemical potential is tuned to the Dirac point. Superconductivity is induced on the surface state by the proximity effect\cite{FuKane}, represented by an extra term $\Delta \psi^\dag_{\uparrow}\psi^\dag_{\downarrow}+\mbox{h.c.}$ to the Hamiltonian. The superconducting layer of height $h_z$ is deposited as described in Figure \ref{fig:long-junc}. At the thin insulating layer between the two annuli, as well as in the inner and outer holes, an insulating magnetic material  of constant magnetization should be deposited, resulting in a Zeeman term of the form $M\psi^\dag \sigma_z \psi$ in the surface Hamiltonian. This breaks the time reversal symmetry at the edges and by that chooses a single chirality for the flow of the neutral chiral edge modes. The dynamics of the soliton will be largely determined  by the $s$-wave superconducting layer, while a zero energy Majorana mode will be trapped by the soliton on the surface state of the topological insulator. Other than these material specific details, the rest of the arguments in the paper can be applied with no further changes.

To estimate the experimental parameters we take typical values $\lambda_L=0.1\mu$m and $\lambda_J=30\mu$m for an $s$-wave superconducting layer with $h_z=0.5\mu$m and an insulating region of width $d=20\AA$. For these parameters, the dimensionless parameter $\beta^2\sim 0.01$, and the velocity of light is $\bar{c}=0.1 c$. We take the neutral edge velocity to be $v_\psi=10^5$m$/$s \cite{FuKane} (which coincides with the surface state Fermi velocity close to the Dirac point), and the tunneling between Majorana edge states to be $0.025$meV. The plasmon gap is $E_p\sim 5$K \cite{HermonShnirmanBenJacob}, while the intra-core gap is given by direct substitution to equation (\ref{eq:spectrum-soliton}), $E_n\sim 120$mK. The junction charging energy is $E_c=(2e)^2/2C\sim 250(\lambda_J/L)^2$mK. In order to observe quantum phenomena, we need the temperature to be smaller than all these energy scales, $T<\min\left\{E_n,E_c,E_p\right\}$. The operating temperatures are therefore in the range of $10-100\,$mK.

The implementation of the Josephson junction in a hybrid structure of semiconductor quantum well coupled to an $s$-wave superconductor and a ferromagnetic insulator\cite{SauLutchynTewari} is similarly straightforward. The superconducting layer is deposited as in Figure \ref{fig:long-junc}, while the ferromagnetic insulator layer will be deposited throughout with no restrictions, breaking the time-reversal symmetry for the quantum well. Another option is to use a quantum well with both Rashba and Dresselhaus spin-orbit coupling, and deposit the superconducting layer as before, with an external magnetic field applied in the in-plane direction \cite{alicea}. Both cases will result in an SNS type Josephson junction.

The implementation using SRO is different in several respects. First, SRO is not spin polarized. As a consequence, in bulk SRO Majorana modes are carried by half-vortices, that are  made of a $\pi$ rotation of the pairing $d$-vector glued together with a $\pi$ phase winding of the order parameter. These half-vortices were recently observed \cite{Budakian} in mesoscopic samples of the type that may be useful for the present context. Similarly, Majorana modes in a long Josephson junction would require half-fluxons.
Second, SRO is a three dimensional material, made of two dimensional layers. The physics described above for the Majorana modes carried by semi-fluxons decouples into different layers, one Majorana edge per layer. The multitude of Majorana modes does not affect the even $N_v$ case. In the case of odd $N_v$, the factor $\gamma_e\gamma_J$ in $U$ is replaced by $\prod_{i=1}^{L_z}\gamma_{e,i}\gamma_{J,i} $, with the index $i$ numbering the layers, and $L_z$ being the number of layers. The unitary transformation still has two eigenvalues that differ by a minus sign, and a transfer of a single electron facilitates a transition between the two states. Thus, as long as the zero energy states do not split, the multi-layer case will not be different from the single layer one.

Tunnel coupling between the layers may split the Majorana zero energy modes. However, a tunneling term between the $i,j$ layers is of the form
\begin{eqnarray}
	H_{\mathrm{coupling}}=2 i t\int dx\, \psi_i(x)\psi_j(x)\sin\left[(\phi_i-\phi_j)/2\right],
\end{eqnarray}
with $i,j$ being the layers involved in the tunneling.
This tunneling is suppressed when the phase difference between the layers vanishes.
Fluctuations of $\phi$ are massive due to the presence of the inter-layer Josephson coupling, and thus we may expect the Majorana modes not to split (see also \cite{Volovik2003}).

Third, SRO may take two possible forms of p-wave pairing, commonly denoted by $p_x\pm ip_y$. The analysis above assumes that both super-conductors are of the same pairing form. If the converse is true, the shape of the Majorana zero mode is different, but the rest of the analysis is unaffected. We analyze this case in the supplementary.

{ We note that SRO suffers} from the presence of a small mini-gap $\sim \Delta^2/E_F$ for quasi-particle excitations in the core of bulk vortices and in edge states. This sets a constraint of about $10$mK on the maximum temperature. For the case of a topological insulator tuned to the Dirac point this problem is avoided.


\section{Summary}

In summary, we predict that phase solitons in a long Josephson junction embedded in a topological superconductor would carry a localized Majorana zero mode. Consequently, these solitons would constitute anyons with non-Abelian exchange statistics. Exploiting the quantum nature of these solitons we suggest two experiments that can reveal the presence of the Majorana modes. One experiment involves voltage oscillations of a topological superconducting capacitor, realized by a circular Josephson junction hosting a single fluxon. The other experiment involves interference effects of a fluxon beam.





\begin{acknowledgments}
We would like to thank P. Bonderson, E. Fradkin, M. Freedman, A. Ludwig, R. Lutchyn, B. Seradjeh and S. Vishveshwara for useful discussions. We are grateful for the hospitality of the Aspen center for physics during which part of this work was carried out. EG would like to thank the ICMT fellowship program for support. AS acknowledges support from the US-Israel Binational Science Foundation, Minerva Foundation and Microsoft Corporation.
\end{acknowledgments}

\section{Supplemental Information}

\subsection{Electron tunneling between Majorana edge states}

In this section we derive the low energy form of the electron tunneling term across a Josephson junction made of two topological superconductors separated by an insulator. We take the topological super-conductors to be $p_x\pm i p_y$ spin-polarized superconductors in the weak pairing phase, where $\mu>0$ \cite{ReadGreen}. We take the insulator to be a region where $\mu<0$, and find it convenient to allow for a pairing term also in the insulating region. We do not expect the physics to be altered much by the small occupations induced by the pairing term in the conduction band of the insulator. Also, we do not expect the results to depend on the microscopic details -- they should apply for any topological superconductor in the same universality class. As we describe below, in the absence of electron tunneling across the insulator the model predicts the presence of two spatially separated Majorana edge modes with a tunnel barrier between them. The modes are counter-propagating for a junction separating two superconductors of the $p_x+ip_y$ type, and co-propagating for a junction separating a $p_x+ip_y$ super-conductor from a $p_x-ip_y$ one.

We start by reviewing the case of a single straight infinite edge state, worked out in details in \cite{ReadGreen,FendleyFisherNayak,RoyStone}. The Hamiltonian density of a $p_x+i p_y$ superconductor in the presence of an edge going along the $x$-axis ($y=0$) is $C^\dag H C/2$, where $C=(c,c^\dag)^T$, $c$ is the electron operator, and
\begin{eqnarray}
	H=\left(\begin{array}{cc} H_0 & \Delta \\ \Delta^\dag & -H_0^T\end{array}\right),
\end{eqnarray}
with
\begin{eqnarray}
	&&H_0=-\frac{1}{2m}\boldsymbol{\nabla}^2+V(y)-\mu,\\
	&&\Delta=i v_\Delta e^{i\phi({\bf r})/2}(\partial_x-i \partial_y)e^{i\phi({\bf r})/2}.
\end{eqnarray}
Note the following symmetry of the Hamiltonian, $\sigma_x H^* \sigma_x=-H$, having the consequence that for any solution $\chi_E=(u_E,v_E)^T$ with energy $E$ there exists a solution $\chi_{-E}=\sigma_x \chi^*_E$ with energy $-E$. Define $\tilde{\mu}(y)=\mu-V(y)$ the electro-chemical potential, which we take to be negative on one side of the edge (the insulator), and positive on the other (the topological super-conductor). Linearizing the Hamitonian near the edge, we get for $H'=e^{-\frac{i}{2}\sigma_z\phi({\bf r})}H e^{\frac{i}{2}\sigma_z \phi({\bf r})}$,
\begin{eqnarray}
	\label{eq:linearized-H}
	H'=-\sigma_z (\tilde{\mu}(y)-v_\Delta\sigma_x\partial_y)+i v_\Delta\sigma_x\partial_x.
\end{eqnarray}
The Hamiltonian factorizes into $x$ and $y$ components. The part that depends on $y$ is an effective one-dimensional Dirac Hamiltonian with a mass term that changes sign at the edge, and therefore contains a zero mode. We assume that the potential is sharp enough so that the higher energy states can be disregarded. We can now write the Nambu spinor projected into this zero energy state, $C_e({\bf r})$ (with the subscript standing for "edge") as
\begin{eqnarray}
	C_e({\bf r})= e^{\frac{i}{2}\sigma_z\phi({\bf r})}e^{\frac{\sigma_x}{v_\Delta}\int_{0}^y \tilde{\mu}(y')dy'}\left(\begin{array}{c}1 \\ \pm 1 \end{array}\right)\psi(x),
\end{eqnarray}
where the effective Hamiltonian for the Hermitian field $\psi(x)$ is
\begin{eqnarray}
	\label{eq:ham-majorana}
	H''=\pm i v_\Delta \partial_x,
\end{eqnarray}
and $\pm$ is the eigenvalue of $\sigma_x$. As is apparent from equation (\ref{eq:linearized-H}), the eigenvalue of $\sigma_x$ is uniquely chosen by the behavior of the electro-chemical potential $\tilde{\mu}$ at $y\to\pm \infty$ to make the zero-mode in the $y$-direction normalisable. This effectively locks the direction of the propagation of the edge to the spatial profile of $\sgn(\tilde{\mu}(y))$. The operator $\psi(x)$ can now be expanded as
\begin{eqnarray}
	\psi(x)=\frac{1}{\sqrt{L}}\sum_k e^{i k x}\psi_k,
\end{eqnarray}
where $\psi_k$ satisfies $\psi_k^\dag=\psi_{-k}$ and $L$ is the length of the junction. 
The energy is related to $k$ via $E=\pm v_\Delta k$, identifying $v_\psi=v_\Delta$.

It will prove useful to generalize the result to a straight edge forming an angle $\theta$ with the $x$-axis \cite{RoyStone,FendleyFisherNayak}, for two reasons. First, it gives a handle on the direction of propagation of the edge, including inverting its chirality. Second, when we consider a circular edge, we can locally describe the edge as being flat, with the parameter $\theta$ affecting the boundary conditions through an explicit monodromy. To proceed, we denote by $X$ the direction along the edge, so that $(x,y)=X (\cos\theta,\sin\theta)$, and define $Y$ as the perpendicular direction,
\begin{eqnarray}
	\nonumber C_e({\bf R})=e^{\frac{i}{2}\sigma_z\phi({\bf R})}e^{-\frac{i}{2}\sigma_z\theta}e^{\frac{\sigma_x}{v_\Delta}\int_{0}^{Y} \tilde{\mu}(Y')dY'}\left(\begin{array}{c}1 \\ \pm 1 \end{array}\right)\psi (X).\\
\end{eqnarray}
where ${\bf R}=(X,Y)$. Note that a rotation by an angle $\theta=\pi$ reverses the chirality of the edge, connecting the two orthogonal chirality states represented by eigenvectors of $\sigma_x$. Also, for a large circular edge we can treat the term which depends on $\theta$ as a rotating ``frame'', leading to anti-periodic boundary conditions. Therefore, when an even number of vortices is enclosed by the edge $k$ is quantized to be $k=\frac{2\pi}{L}(n+1/2)$ ($n\in\mathbb{Z}$), while for an odd number of vortices $k=\frac{2\pi}{L}n$ , so that the spectrum contains a zero mode \cite{ReadGreen}.

Suppose we have two edges, at $y=\pm a/2$, where $\tilde{\mu}(y)<0$ for $-a/2<y<a/2$ and $\tilde{\mu}(y)>0$ otherwise. Assume the form of the order parameter is $p_x+i p_y$ throughout the sample. We can write the Nambu spinor projected on the two edge states as
 \begin{eqnarray}
		\nonumber &&C_{e,1}=
e^{-\frac{1}{v_\Delta}\int_{a/2}^y \tilde{\mu}(y')dy'}\left(\begin{array}{c}e^{-\frac{i\pi}{2}+\frac{i\phi_1}{2}} \\ e^{\frac{i\pi}{2}-\frac{i\phi_1}{2}}\end{array}\right)\psi_1(x),\\
		&&C_{e,2}= 
e^{\frac{1}{v_\Delta}\int_{-a/2}^y \tilde{\mu}(y')dy'}\left(\begin{array}{c}e^{\frac{i\phi_2}{2}} \\ e^{-\frac{i\phi_2}{2}}\end{array}\right)\psi_2(x).
\end{eqnarray}
Hence, electron tunneling between edges is of the form
\begin{eqnarray}
	&&=i m\int dx\,\psi_1(x)\psi_2(x) \cos\left[\frac{\phi_1(x)-\phi_2(x)}{2}\right].
\end{eqnarray}

In the case that the insulator separates a $p_x+i p_y$ superconductor from a $p_x-i p_y$ superconductor, for the $p_x-i p_y$ sector there will be an extra minus sign preceding the term containing $\partial_y$ in equation (\ref{eq:linearized-H}). This would flip again the eigenvalue of $\sigma_x$ with respect to the orientation, so that $C_{e,1}$ will be redefined to be
 \begin{eqnarray}
		C_{e,1}= 
e^{-\frac{1}{v_\Delta}\int_{a/2}^y \tilde{\mu}(y')dy'}\left(\begin{array}{c}e^{\frac{i\phi_1}{2}} \\ e^{-\frac{i\phi_1}{2}}\end{array}\right)\psi_1(x),
\end{eqnarray}
leading to the following form for the tunneling term
\begin{eqnarray}
	&&=i m\int dx\,\psi_1(x)\psi_2(x) \sin\left[\frac{\phi_1(x)-\phi_2(x)}{2}\right].
\end{eqnarray}

\subsection{Solutions for the Majorana mode for different cases}

In this section we find zero energy solutions of the neutral sector of our main Hamiltonian. Written in a general form, the Hamiltonian is
\begin{eqnarray}
	H=\left[\begin{array}{cc} i v_\psi\partial_x & i W_\pm (x)\\ -i W_\pm(x) & \pm i v_\psi \partial_x \end{array}\right].
\end{eqnarray}
where $W_+(x)=m\sin(\phi/2)$, $W_-(x)=m\cos(\phi/2)$, and $\pm$ denote the relative chirality of the two edge states. To solve for zero modes, we write $H\chi=0$ where $\chi=(f,g)^T$. Squaring the Hamiltonian, we get
\begin{eqnarray}
	\label{eq:diff-eq} \left[\partial_{\tilde{x}}^2-\frac{\partial_{\tilde{x}} W_\pm(\tilde{x})}{W_\pm(\tilde{x})}\partial_{\tilde{x}} \pm \frac{\eta^2}{m^2} W_\pm^2(\tilde{x})\right]g(\tilde{x})=0
\end{eqnarray}
Here $\eta=\frac{m \ell}{v_\psi}$ with $\ell$ a length scale related to the potential $W_\pm(x)$, and $\tilde{x}=x/\ell$. A general solution for co-propagating Majorana states is (define $\Phi_\pm(\tilde{x})=\frac{\eta}{m} \int^{\tilde{x}} W_\pm(\tilde{x}')d\tilde{x}'$)
\begin{eqnarray}
	\nonumber &&f_+(\tilde{x})=-A \sin\left[\Phi_+(\tilde{x})\right]+B \cos\left[\Phi_+(\tilde{x})\right],\\
	\label{eq:maj-plus} &&g_+(\tilde{x})=A \cos\left[\Phi_+(\tilde{x})\right]+B \sin\left[\Phi_+(\tilde{x})\right].
\end{eqnarray}
For counter-propagating Majorana states one gets
\begin{eqnarray}
	\nonumber  &&f_-(\tilde{x})=-A \sinh\left[\Phi_-(\tilde{x})\right]-B \cosh\left[\Phi_-(\tilde{x})\right],\\
	\label{eq:maj-minus} &&g_-(\tilde{x})=A \cosh\left[\Phi_-(\tilde{x})\right]+B \sinh\left[\Phi_-(\tilde{x})\right].
\end{eqnarray}
The resulting Majorana fermion acquires the form,
\begin{eqnarray}
	\gamma_{J,\pm}=\int dx \left[f_\pm(x)\psi_1(x)+g_\pm(x)\psi_2(x)\right].
\end{eqnarray}

In the body of the paper we derive the zero energy modes for the case of a long Josephson junction separating two $p_x+ip_y$ super-conductors, where the two Majorana modes propagate in opposite directions.  
In the following, we shall write down explicitly the solutions for two other case, that of co-moving Majorana modes, and that of a short Josephson junction. 

\subsubsection{Co-propagating Majorana edge states}

\begin{figure}[htp]
\centering
\includegraphics[scale=0.28]{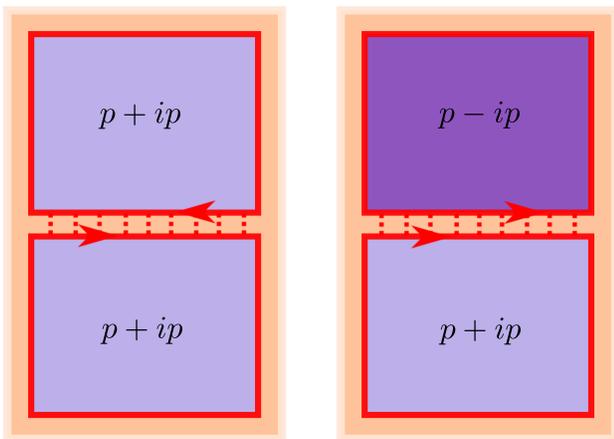}
\caption{A $p_x+i p_y\, / \, p_x+ip_y$ junction versus a $p_x+i p_y\, / \, p_x-i p_y$ junction. The direction of the propagation of the Majorana edge states is marked by arrows. Dotted lines indicate electron tunneling.}
\label{fig:supp-pip}
\end{figure}

The Hamiltonian of the neutral modes for a $p_x+i p_y\,/\,p_x-i p_y$ junction is different from a $p_x+i p_y\,/\,p_x+i p_y$ junction for two reasons: First, the Majorana edge states are propagating in the same direction, and second, the $\cos(\phi/2)$ in the mass term is replaced with $\sin(\phi/2)$, with $\phi$ the phase difference between the two superconductors. The resulting Hamiltonian in the presence of a soliton is
\begin{eqnarray}
	H=\left[\begin{array}{cc} i v_\psi\partial_x & i m \,\mbox{sech}\left(\frac{x}{\lambda_J}\right)\\ -i m \,\mbox{sech}\left(\frac{x}{\lambda_J}\right) & i v_\psi \partial_x \end{array}\right].
\end{eqnarray}
For a closed long circular junction the zero energy solution to that equation is of the class presented in equation (\ref{eq:maj-plus}), with
\begin{eqnarray}
	\Phi_+(x)=2 \eta \arctan\left[\tanh\left(\frac{x}{2\lambda_J}\right)\right],
\end{eqnarray}
where $\eta=m \lambda_J/v_\psi$ and $-\frac{L}{2}<x<\frac{L}{2}$. The two functions $f_+$ and $g_+$ should also satisfy the boundary conditions imposed by the magnetic flux configuration. For an odd number of vortices within the central hole $f$ is periodic while $g$ should be anti-periodic, hence $A=0$. Similarly, for an even number of vortices $B=0$ (In the absence of a fluxon both $f$ and $g$ have the same boundary condition, hence both $A$ and $B$ must vanish, and there is no zero energy solution). In contrast to the case of counter-propagating edge modes, where the Majorana state is localized close to the center of the soliton, in the case of co-propagating edge modes it is delocalized throughout the junction. However, the profile of the fluxon imposes a sign change on the part of the Majorana mode's wave-function which is bound to the anti-periodic edge. Hence, when the fluxon is adiabatically carried around the junction, the boundary conditions result in the usual transformation applied to this delocalized Majorana mode, i.e. $\gamma_J\to (-1)^{N_v}\gamma_J$.

\subsubsection{The case of a large soliton}

In the main text we consider the case that $\lambda_J\ll L$ and $\frac{v_\psi}{m}\ll \lambda_J$. Here we will discuss the solution in the limit $\lambda_J\gg L$ (a short Josephson junction). In this limit, the phase interpolates linearly between $0$ and $2\pi$, so we can write $\phi=\frac{2\pi}{L}x+\pi$ (taking the center of the soliton to be at $x=0$). For a junction of counter-propagating edge modes, the neutral sector of the Hamiltonian takes the approximate form
\begin{eqnarray}
	H=\left[\begin{array}{cc} i v_\psi\partial_x & -i m \sin\left(\frac{\pi x}{L}\right)\\ i m \sin\left(\frac{\pi x}{L}\right)& -i v_\psi \partial_x \end{array}\right],
\end{eqnarray}
where $\eta=m L/v_\psi$. The solution is of the class presented in equation (\ref{eq:maj-minus}), with
\begin{eqnarray}
	\Phi_-(x)=\frac{\eta}{\pi}\cos\left(\frac{\pi x}{L}\right),
\end{eqnarray}
where we have dropped a constant term by redefining $A$ and $B$ in equation (\ref{eq:maj-minus}). Most of the weight of the wave-function is therefore present in the vicinity of the soliton. 

For a short junction where the edge modes are co-propagating the solution is of the class presented in equation (\ref{eq:maj-plus}), with
\begin{eqnarray}
	\label{eq:phi-plus} \Phi_+(x)=\frac{\eta}{\pi}\sin\left(\frac{\pi x}{L}\right).
\end{eqnarray}
For this case, the Majorana solution is uniformly spread throughout the junction, with its weight shifting between the inner and outer edges to conform with the boundary conditions. Here it is straightforward to see explicitly the effects of the boundary conditions. When $x$ is increased by $L$, we get that $\Phi_\pm(x+L)=-\Phi_\pm(x)$ and, consequently, the odd functions ($\sin$, $\sinh$) in equations (\ref{eq:maj-plus})-(\ref{eq:maj-minus}) change sign, while the even functions ($\cos$, $\cosh$) stay the same. Due to the boundary conditions imposed on $f$ and $g$, one periodic while the other anti-periodic, either $A=0$ or $B=0$. When combined with $\phi_1\to\phi_1+2\pi$ (originating from the effect of the fluxon on the phase of the superconductor enclosed by its path, and leading to $\psi_1 \to -\psi_1$), this would result in the usual transformation rule for the Majorana operator defined by this solution, i.e. $\gamma_J\to (-1)^{N_v}\gamma_J$.

In summary, independent of the size of the soliton relative to the size of the junction and of the relative direction of motion of the two edge modes within the junction, a winding of the soliton around the junction results in the transformation $\gamma_J\to (-1)^{N_v}\gamma_J$. Thus, at least to the extent that we deal with a single fluxon in the junction, the effects associated with the Majorana zero energy state induced by the fluxon in the junction are independent of these details. 

\subsection{Classical fluxon propagated by a super-current}

In this section we consider the bound states on a classical fluxon moving at velocity $v_\phi$. We assume that the junction is oriented along the $x$-direction. To move the fluxon, we apply a super-current along the $y$-direction. For low values of the applied current $I$ the attained velocity of the fluxon is linear in $I$ with a coefficient that depends on the microscopic details of the junction \cite{McLaughlinScott}. When the fluxon moves at a velocity $v_\phi$, it generates a voltage along the perpendicular direction given by
\begin{eqnarray}
	V=\phi_0 v_\phi/L,
\end{eqnarray}
where $L$ is the length of the junction and $\phi_0=h/2e$ is the superconductor flux quantum. This behavior generates the typical $I-V$ characteristics of the junction.

The Hamiltonian for the neutral states in the background of a moving soliton is given by
\begin{eqnarray}
	i\partial_t \Psi_\pm=\left[\begin{array}{cc} i v_\psi \partial_x & i W_\pm (x-v_\phi t) \\ -i W_\pm(x-v_\phi t) & \pm i v_\psi \partial_x \end{array}\right]\Psi_\pm,
\end{eqnarray}
where $\Psi=(\psi_1,\psi_2)^T$. We solve this equation by going to the frame which moves with the soliton. Defining $x'=x-v_\phi t$ to be our new coordinate,
\begin{eqnarray}
	\nonumber && \partial_t=\partial_{t'}-v_\phi\partial_{x'},\\
	&& \partial_x=\partial_{x'},
\end{eqnarray}
we consequently get
\begin{eqnarray}
	i\partial_{t'}\Psi'_\pm=i v_\phi\partial_{x'}\Psi'_\pm+\left[\begin{array}{cc} i v_\psi \partial_{x'} & i W_\pm(x') \\ -i W_\pm(x') & \pm i v_\psi \partial_{x'} \end{array}\right]\Psi'_\pm.
\end{eqnarray}
For co-propagating Majorana edge states the solution is trivial, since $v_\psi$ is simply replaced by $v_\psi+v_\phi$. The more interesting case is when the Majorana edge states are counter-propagating, and the Majorana fermion is localized. Then, when solving for the zero mode, we can square the Hamiltonian and get the result appearing in the main text for the form of the zero energy wave-function, but with the parameter $\eta=m \lambda_J/v_\psi$ replaced by $\eta/\sqrt{1-\left(v_\phi/v_\psi\right)^2}$. The size of the Majorana mode wave-function therefore decreases with increasing speed of the soliton, with a Lorentz-like dependence on its velocity and an effective speed of light which is set to the Majorana edge velocity $v_\psi$ (the soliton decreases in size as well, with the effective speed of light being $\bar{c}\gg v_\psi$).


\begin{thebibliography}{10}

\expandafter\ifx\csname url\endcsname\relax
  \def\url#1{\texttt{#1}}\fi
\expandafter\ifx\csname urlprefix\endcsname\relax\def\urlprefix{URL }\fi
\providecommand{\bibinfo}[2]{#2}
\providecommand{\eprint}[2][]{\url{#2}}

\bibitem{MooreRead}
\bibinfo{author}{{Moore}, G.} \& \bibinfo{author}{{Read}, N.}
\newblock \bibinfo{title}{{Nonabelions in the fractional quantum hall effect}}.
\newblock \emph{\bibinfo{journal}{Nuclear Physics B}}
  \textbf{\bibinfo{volume}{360}}, \bibinfo{pages}{362--396}
  (\bibinfo{year}{1991}).

\bibitem{NayakWilczek}
\bibinfo{author}{{Nayak}, C.} \& \bibinfo{author}{{Wilczek}, F.}
\newblock \bibinfo{title}{{2n-quasihole states realize $2^{n-1}$-dimensional
  spinor braiding statistics in paired quantum Hall states}}.
\newblock \emph{\bibinfo{journal}{Nuclear Physics B}}
  \textbf{\bibinfo{volume}{479}}, \bibinfo{pages}{529--553}
  (\bibinfo{year}{1996}).
\newblock \eprint{arXiv:cond-mat/9605145}.

\bibitem{RMP}
\bibinfo{author}{{Nayak}, C.}, \bibinfo{author}{{Simon}, S.~H.},
  \bibinfo{author}{{Stern}, A.}, \bibinfo{author}{{Freedman}, M.} \&
  \bibinfo{author}{{Das Sarma}, S.}
\newblock \bibinfo{title}{{Non-Abelian anyons and topological quantum
  computation}}.
\newblock \emph{\bibinfo{journal}{Rev. Mod. Phys.}} \textbf{\bibinfo{volume}{80}},
  \bibinfo{pages}{1083--1159} (\bibinfo{year}{2008}).

\bibitem{Kitaev}
\bibinfo{author}{{Kitaev}, A.~Y.}
\newblock \bibinfo{title}{{Fault-tolerant quantum computation by anyons}}.
\newblock \emph{\bibinfo{journal}{Ann. of Phys.}}
  \textbf{\bibinfo{volume}{303}}, \bibinfo{pages}{2--30}
  (\bibinfo{year}{2003}).

\bibitem{DasSarmaNayakFreedman}
\bibinfo{author}{Das~Sarma, S.}, \bibinfo{author}{Freedman, M.} \&
  \bibinfo{author}{Nayak, C.}
\newblock \bibinfo{title}{Topologically protected qubits from a possible
  non-abelian fractional quantum hall state}.
\newblock \emph{\bibinfo{journal}{Phys. Rev. Lett.}}
  \textbf{\bibinfo{volume}{94}}, \bibinfo{pages}{166802}
  (\bibinfo{year}{2005}).

\bibitem{FreedmanLarsenWang}
\bibinfo{author}{Freedman, M.~H.}, \bibinfo{author}{Kitaev, A.},
  \bibinfo{author}{Larsen, M.~J.} \& \bibinfo{author}{Wang, Z.}
\newblock \bibinfo{title}{Topological quantum computation}.
\newblock \emph{\bibinfo{journal}{Bull. Amer. Math. Soc.}}
  \textbf{\bibinfo{volume}{40}}, \bibinfo{pages}{31--38}
  (\bibinfo{year}{2003}).

\bibitem{SternHalperin}
\bibinfo{author}{Stern, A.} \& \bibinfo{author}{Halperin, B.~I.}
\newblock \bibinfo{title}{Proposed experiments to probe the non-abelian
  $\nu{}=5/2$ quantum hall state}.
\newblock \emph{\bibinfo{journal}{Phys. Rev. Lett.}}
  \textbf{\bibinfo{volume}{96}}, \bibinfo{pages}{016802}
  (\bibinfo{year}{2006}).

\bibitem{BondersonKitaevShtengel}
\bibinfo{author}{Bonderson, P.}, \bibinfo{author}{Kitaev, A.} \&
  \bibinfo{author}{Shtengel, K.}
\newblock \bibinfo{title}{Detecting non-abelian statistics in the $\nu{}=5/2$
  fractional quantum hall state}.
\newblock \emph{\bibinfo{journal}{Phys. Rev. Lett.}}
  \textbf{\bibinfo{volume}{96}}, \bibinfo{pages}{016803}
  (\bibinfo{year}{2006}).

\bibitem{KopninSalomaa}
\bibinfo{author}{Kopnin, N.~B.} \& \bibinfo{author}{Salomaa, M.~M.}
\newblock \bibinfo{title}{{Mutual friction in superfluid He-3: Effects of bound
  states in the vortex core}}.
\newblock \emph{\bibinfo{journal}{Phys. Rev. B}} \textbf{\bibinfo{volume}{44}},
  \bibinfo{pages}{9667--9677} (\bibinfo{year}{1991}).

\bibitem{ReadGreen}
\bibinfo{author}{Read, N.} \& \bibinfo{author}{Green, D.}
\newblock \bibinfo{title}{Paired states of fermions in two dimensions with
  breaking of parity and time-reversal symmetries and the fractional quantum
  hall effect}.
\newblock \emph{\bibinfo{journal}{Phys. Rev. B}} \textbf{\bibinfo{volume}{61}},
  \bibinfo{pages}{10267--10297} (\bibinfo{year}{2000}).

\bibitem{Ivanov}
\bibinfo{author}{Ivanov, D.~A.}
\newblock \bibinfo{title}{Non-abelian statistics of half-quantum vortices in
  $p$-wave superconductors}.
\newblock \emph{\bibinfo{journal}{Phys. Rev. Lett.}}
  \textbf{\bibinfo{volume}{86}}, \bibinfo{pages}{268--271}
  (\bibinfo{year}{2001}).

\bibitem{ReznikAharonov}
\bibinfo{author}{{Reznik}, B.} \& \bibinfo{author}{{Aharonov}, Y.}
\newblock \bibinfo{title}{{Question of the nonlocality of the Aharonov-Casher
  effect}}.
\newblock \emph{\bibinfo{journal}{Phys. Rev. D}} \textbf{\bibinfo{volume}{40}},
  \bibinfo{pages}{4178--4183} (\bibinfo{year}{1989}).

\bibitem{Wees}
\bibinfo{author}{{van Wees}, B.~J.}
\newblock \bibinfo{title}{{Aharonov-Bohm-type effect for vortices in
  Josephson-junction arrays}}.
\newblock \emph{\bibinfo{journal}{Phys. Rev. Lett.}} \textbf{\bibinfo{volume}{65}},
  \bibinfo{pages}{255--258} (\bibinfo{year}{1990}).

\bibitem{HermonSternBenJacob}
\bibinfo{author}{{Hermon}, Z.}, \bibinfo{author}{{Stern}, A.} \&
  \bibinfo{author}{{Ben-Jacob}, E.}
\newblock \bibinfo{title}{{Quantum dynamics of a fluxon in a long circular
  Josephson junction}}.
\newblock \emph{\bibinfo{journal}{Phys. Rev. B}} \textbf{\bibinfo{volume}{49}},
  \bibinfo{pages}{9757--9762} (\bibinfo{year}{1994}).

\bibitem{Elion}
\bibinfo{author}{{Elion}, W.~J.}, \bibinfo{author}{{Wachters}, J.~J.},
  \bibinfo{author}{{Sohn}, L.~L.} \& \bibinfo{author}{{Mooij}, J.~E.}
\newblock \bibinfo{title}{{The Aharonov-Casher effect for vortices in
  Josephson-junction arrays}}.
\newblock \emph{\bibinfo{journal}{Phys. B Cond. Mat.}}
  \textbf{\bibinfo{volume}{203}}, \bibinfo{pages}{497--503}
  (\bibinfo{year}{1994}).

\bibitem{AharonovCasher}
\bibinfo{author}{Aharonov, Y.} \& \bibinfo{author}{Casher, A.}
\newblock \bibinfo{title}{Topological quantum effects for neutral particles}.
\newblock \emph{\bibinfo{journal}{Phys. Rev. Lett.}}
  \textbf{\bibinfo{volume}{53}}, \bibinfo{pages}{319--321}
  (\bibinfo{year}{1984}).

\bibitem{GrosfeldSeradjehVishveshwara}
\bibinfo{author}{{Grosfeld}, E.}, \bibinfo{author}{{Seradjeh}, B.} \&
  \bibinfo{author}{{Vishveshwara}, S.}
\newblock \bibinfo{title}{{Proposed Aharonov-Casher interference measurement of non-Abelian vortices in chiral $p$-wave superconductors}}.
\newblock \emph{\bibinfo{journal}{{Phys. Rev. B}}}
  \textbf{\bibinfo{volume}{83}}, \bibinfo{pages}{104513}
(\bibinfo{year}{2011}).

\bibitem{Tinkham}
\bibinfo{author}{{Tinkham}, M.}
\newblock \emph{\bibinfo{title}{Introduction to Superconductivity}}
  (\bibinfo{publisher}{Dover Publications}, \bibinfo{address}{Mineola, New
  York}, \bibinfo{year}{2004}), \bibinfo{edition}{2nd} edn.

\bibitem{IlanGrosfeldStern}
\bibinfo{author}{{Ilan}, R.}, \bibinfo{author}{{Grosfeld}, E.} \& \bibinfo{author}{{Stern}, A.}
\newblock \bibinfo{title}{{Coulomb Blockade as a Probe for Non-Abelian Statistics in Read-Rezayi States}}.
\newblock \emph{\bibinfo{journal}{Phys. Rev. Lett.}} \textbf{\bibinfo{volume}{100}},
  \bibinfo{pages}{086803} (\bibinfo{year}{2008}).

\bibitem{Liang}
\bibinfo{author}{{Liang}, F.}
\newblock \bibinfo{title}{{Electron Teleportation via Majorana Bound States in a Mesoscopic Superconductor}}.
\newblock \emph{\bibinfo{journal}{Phys. Rev. Lett.}} \textbf{\bibinfo{volume}{104}},
  \bibinfo{pages}{056402} (\bibinfo{year}{2010}).

\bibitem{ShnirmanBenJacobMalomed}
\bibinfo{author}{{Shnirman}, A.}, \bibinfo{author}{{Ben-Jacob}, E.} \&
  \bibinfo{author}{{Malomed}, B.}
\newblock \bibinfo{title}{{Tunneling and resonant tunneling of fluxons in a
  long Josephson junction}}.
\newblock \emph{\bibinfo{journal}{Phys. Rev. B}} \textbf{\bibinfo{volume}{56}},
  \bibinfo{pages}{14677--14685} (\bibinfo{year}{1997}).

\bibitem{Wallraff}
\bibinfo{author}{{Wallraff}, A.} \emph{et~al.}
\newblock \bibinfo{title}{{Quantum dynamics of a single vortex}}.
\newblock \emph{\bibinfo{journal}{Nature}} \textbf{\bibinfo{volume}{425}},
  \bibinfo{pages}{155--158} (\bibinfo{year}{2003}).

\bibitem{Hassler}
\bibinfo{author}{{Hassler}, F.}, \bibinfo{author}{{Akhmerov}, A.~R.}, \bibinfo{author}{{Hou}, C-Y} \& \bibinfo{author}{{Beenakker}, C.~W.~J.}
\newblock \bibinfo{title}{{Anyonic interferometry without anyons: how a flux qubit can read out a topological qubit}}.
\newblock \emph{\bibinfo{journal}{New J. Phys.}} \textbf{\bibinfo{volume}{12}},
  \bibinfo{pages}{125002} (\bibinfo{year}{2010}).

\bibitem{ClarkeShtengel}
\bibinfo{author}{{Clarke}, D.~J.} \& \bibinfo{author}{{Shtengel}, K.}
\newblock \bibinfo{title}{{Improved phase-gate reliability in systems with neutral Ising anyons}}.
\newblock \emph{\bibinfo{journal}{Phys. Rev. B}} \textbf{\bibinfo{volume}{82}},
  \bibinfo{pages}{180519} (\bibinfo{year}{2010}).

\bibitem{HermonShnirmanBenJacob}
\bibinfo{author}{Hermon, Z.}, \bibinfo{author}{Shnirman, A.} \&
  \bibinfo{author}{Ben-Jacob, E.}
\newblock \bibinfo{title}{Dephasing length and coherence of a quantum soliton
  in an ideal long josephson junction}.
\newblock \emph{\bibinfo{journal}{Phys. Rev. Lett.}}
  \textbf{\bibinfo{volume}{74}}, \bibinfo{pages}{4915--4918}
  (\bibinfo{year}{1995}).

\bibitem{SauLutchynTewari}
\bibinfo{author}{{Sau}, J.~D.}, \bibinfo{author}{{Lutchyn}, R.~M.},
  \bibinfo{author}{{Tewari}, S.} \& \bibinfo{author}{{Das Sarma}, S.}
\newblock \bibinfo{title}{{Generic new platform for topological quantum
  computation using semiconductor heterostructures}}.
\newblock \emph{\bibinfo{journal}{Phys. Rev. Lett.}} \textbf{\bibinfo{volume}{104}},
  \bibinfo{pages}{040502} (\bibinfo{year}{2010}).

\bibitem{FuKane}
\bibinfo{author}{{Fu}, L.} \& \bibinfo{author}{{Kane}, C.~L.}
\newblock \bibinfo{title}{{Superconducting proximity effect and majorana
  fermions at the surface of a topological insulator}}.
\newblock \emph{\bibinfo{journal}{Phys. Rev. Lett.}} \textbf{\bibinfo{volume}{100}},
  \bibinfo{pages}{096407} (\bibinfo{year}{2008}).

\bibitem{alicea}
\bibinfo{author}{Alicea, J.}
\newblock \bibinfo{title}{Majorana fermions in a tunable semiconductor device}.
\newblock \emph{\bibinfo{journal}{Phys. Rev. B}} \textbf{\bibinfo{volume}{81}},
  \bibinfo{pages}{125318} (\bibinfo{year}{2010}).

\bibitem{Budakian}
\bibinfo{author}{J. Jang et al.}
\newblock \bibinfo{title}{Observation of Half-Height Magnetization Steps in Sr$_2$RuO$_4$}.
\newblock \emph{\bibinfo{journal}{Science}} \textbf{\bibinfo{volume}{14}},
  \bibinfo{pages}{186-188} (\bibinfo{year}{2011}).

\bibitem{Volovik2003}
\bibinfo{author}{{Volovik}, G.}
\newblock \emph{\bibinfo{title}{The Universe in a Helium Droplet}}
  (\bibinfo{publisher}{Oxford University Press}, \bibinfo{address}{USA}, \bibinfo{year}{2003}).

\bibitem{FendleyFisherNayak}
\bibinfo{author}{Fendley, P.}, \bibinfo{author}{Fisher, M. P.~A.} \&
  \bibinfo{author}{Nayak, C.}
\newblock \bibinfo{title}{Edge states and tunneling of non-abelian
  quasiparticles in the $\nu{}=5/2$ quantum hall state and $p+ip$
  superconductors}.
\newblock \emph{\bibinfo{journal}{Phys. Rev. B}} \textbf{\bibinfo{volume}{75}},
  \bibinfo{pages}{045317} (\bibinfo{year}{2007}).

\bibitem{RoyStone}
\bibinfo{author}{Stone, M.} \& \bibinfo{author}{Roy, R.}
\newblock \bibinfo{title}{Edge modes, edge currents, and gauge invariance in
  $p_{x}+ip_{y}$ superfluids and superconductors}.
\newblock \emph{\bibinfo{journal}{Phys. Rev. B}} \textbf{\bibinfo{volume}{69}},
  \bibinfo{pages}{184511} (\bibinfo{year}{2004}).

\bibitem{McLaughlinScott}
\bibinfo{author}{McLaughlin, D.~W.} \& \bibinfo{author}{Scott, A.~C.}
\newblock \bibinfo{title}{Perturbation analysis of fluxon dynamics}.
\newblock \emph{\bibinfo{journal}{Phys. Rev. A}} \textbf{\bibinfo{volume}{18}},
  \bibinfo{pages}{1652--1680} (\bibinfo{year}{1978}).

\end{thebibliography}
\end{document}